\documentclass[12pt]{article}
\usepackage{epsfig}
\usepackage{amsmath}
\usepackage{graphicx}
\usepackage{latexsym}

\begin{document}

\begin{center}
{\Large \bf Stability of Yang-Mills-Higgs field system\\ in the homogeneous self-dual vacuum field}\\
\vspace{0.5cm}
{\large {\bf V.I. Kuvshinov}}\\
\textsf{Joint Institute for Power and Nuclear Research,\\
  acad. A.K. Krasina 99, Minsk,  220109, Belarus} \\
{\it   E-mail: V.Kuvshinov@sosny.bas-net.by}\\
\vspace{0.5cm}
{\large {\bf V.A. Piatrou}}\\
\textsf{Belarusian State University,\\
 2 Nezalejnasci av., Minsk, 220050, Belarus} \\
{\it  E-mail:     PiatrouVadzim@tut.by}
\end{center}

\begin{abstract}
Classical dynamics of SU(2) model gauge field system with Higgs
field is considered in the homogeneous nonperturbative self-dual
vacuum field. The regions of instability in parametric space are
detected and described analytically.
\end{abstract}

\section*{Introduction}

Much attention has been paid in the last decade to chaos in quantum
field theory. At first, non-abelian Yang-Mills gauge fields were
investigated without spontaneous symmetry breaking. It was shown
analytically and numerically  that classical Yang-Mills theories are
inherently chaotic ones \cite{MST,Savvidy}. Further research of
spatially homogeneous field configurations \cite{SHS} showed that
spontaneous symmetry breakdown leads to order-chaos transition at
some density of energy of classical gauge fields \cite{regular,
SavvidyNucl}, while dynamics of gauge fields in the absence of
spontaneous symmetry breakdown is chaotic at any density of energy.
Study of chaos in classical gauge Yang-Mills theories gives an
opportunity to understand the influence of Higgs fields on chaotic
dynamics of gauge field system. It was demonstrated that classical
Higgs fields regularize chaotic dynamics of classical gauge fields
at densities of energy less than critical and lead to appearance of
order-chaos transition \cite{regular}.  Chaos in Yang-Mills fields
is also considered in connection with confinement \cite{PLA}.

It was shown \cite{Previous} that quantum fluctuations of vector
gauge fields in $SU(2)\bigotimes U(1)$ theory lead to regularization
the dynamics of gauge fields at low densities of energy, and
order-chaos transition occurs with rise of energy density of gauge
fields. It was also observed that if the ratio of coupling constants
of Yang-Mills and Higgs fields is larger than some critical value,
quantum corrections do not affect the chaotic dynamics of Yang-Mills
and Higgs fields. It was demonstrated that centrifugal term in the
model Hamiltonian increases the region of regular dynamics of
Yang-Mills and Higgs fields system at low densities of energy
\cite{Q1}.

The system of Yang-Mills and Higgs fields has an infinite number of
degrees of freedom and it is too complicated to be investigated
directly. In order to reduce the number of degrees of freedom,
following other authors, we consider only spatially homogeneous
fields. This model is a particular case of the general one.
Spatially homogeneous field models allow one to investigate the main
properties of inhomogeneous fields.

In this paper classical dynamics of SU(2) model gauge field system
with Higgs field is considered in the homogeneous self-dual vacuum
field. Various properties of this field in SU(2) gauge theory were
investigated originall in \cite{S, L1, L2, M}. It was demonstrated
that self-dual homogeneous field provides the Wilson confinement
criterion \cite{SU2}. Therefore this field is at least a possible
source of confinement in QCD if it can be shown that such a field is
a dominant configuration in the QCD functional integral \cite{SU2,
QCD}.

Our model consists of homogeneous perturbative and nonperturbative
Yang-Mills field and the Higgs fields. In this paper we investigate
the influence of nonperturbative homogeneous self-dual field on
chaotic dynamics of Yang-Mills and Higgs fields. We demonstrate that
dynamics of our system depends on the parameters of the model.
Stable and chaotic regions in parametric space are detected. Their
bounds are described analytically.

\section{Homogeneous self-dual field}

Homogeneous self-dual field is characterized by the next expressions
\cite{ L1, L2, SU2}:
%\begin{eqnarray}
$$
B^{a}_{\mu}=\frac12 n^{a} B_{\mu \nu} x_{\nu}, \quad B_{\mu
\nu}=-B_{\nu \mu} $$ $$ B_{\mu \nu} B_{\mu \rho}=B^{2} \delta_{\nu
\rho}, \quad B=const $$ $$ \label{dual} \widetilde{B}_{\mu
\nu}=\frac 1 2 \varepsilon_{\mu \nu \alpha \beta} B_{\alpha
\beta}=\pm B_{\mu \nu} $$ $$ B_{i j}=-\epsilon_{i j k} B_{k}, \quad
B_{j 4}=\pm B_{j}. \label{EB}
$$
%\end{eqnarray}
The positive and the negative signs in the last two lines correspond
to the self-dual and anti-self-dual fields respectively. The color
vector $n^{a}$ points in some fixed direction which can be chosen as
$(n^{1},n^{2},n^{3}) =(0,0,1)$ \cite{L1, L2}.

\section{Model potential of the system}

The Lagrangian of SU(2) gauge theory with the Higgs field  in
Euclidean metrics is
%\begin{equation}
$$
L=-\frac{1}{4} G^{a}_{\mu \nu} G^{a}_{\mu
\nu}-\frac{1}{2}(D_{\mu}\phi)^{+}(D_{\mu}\phi)-V(\phi),
$$
%\end{equation}
where $\phi$ is the real triplet of the Higgs scalar fields.
Classical potential of Higgs fields has the following form
%\begin{equation}
$$
V(\phi)=\mu^{2}\phi^{2}+\lambda \phi^{4},
$$
%\end{equation}
The Yang-Mills field tensor is:
%\begin{equation}
$$
\label{Strength}
 G^{a}_{\mu \nu}=\partial_{\mu} A^{a}_{\nu} -
\partial_{\nu} A^{a}_{\mu} +
g \epsilon^{a b c} A^{b}_{\mu} A^{c}_{\nu},
$$
%\end{equation}
and
%\begin{equation}
$$
 \label{CovDif}
D_{\mu}\phi=\partial_{\mu}\phi - i g  A^{b}_{\mu} T^{b} \phi,
$$
%\end{equation}
where $A^{a},\quad a=1,2,3$ are the three non-abelian Yang-Mills
fields and $T^{a}$ are the generators of the group SU(2) in adjoint
representation, $g$ denotes the coupling constant of non-abelian
gauge fields.

The vacuum in our model is realized by the homogeneous self-dual
field. This field represents the nonperturbative component of
Yang-Mills field potential. We regard  self-dual field as an
external field. It is taken into account by substituting modified
vector potential in the Yang-Mills-Higgs Lagrangian
%\begin{equation}
$$
A^{a}_{\mu} \rightarrow A^{a}_{\mu}+B^{a}_{\mu}
$$
%\end{equation}
where $A^{a}_{\mu}$ is the perturbative component and  $B^{a}_{\mu}$
is the nonperturbative component of the Yang-Mills field.
%Nonperturbative component $B^{a}_{\mu}$ we regard as an external
%field.

We work in the gauge:
%\begin{equation}
$$
A^{a}_{0}= 0,
$$
%\end{equation}
and consider spatially homogeneous field configurations
\cite{regular}
%\begin{equation}
$$
\partial_{i}A^{a}_{\mu}= 0, \qquad \partial_{i}\phi^{a}= 0, \qquad
i=1..3.
$$
%\end{equation}

When $\mu^{2}<0$ the Higgs potential $V(\phi)$ has a minimum at
non-zero $\phi$:
%\begin{equation}
$$
|\underline{\phi_{0}}|=\sqrt{\frac{-\mu^{2}}{4 \lambda}}=\upsilon,
$$
%\end{equation}
This Higgs vacuum is degenerate and after spontaneous symmetry
breaking it can be chosen as
%\begin{equation}\label{Higgs}
$$
\underline{\phi_{0}}=(\phi_{1},\phi_{2},\phi_{3})=(0,0,\upsilon).
$$
%\end{equation}

The direction of the nonperturbative field can be chosen
arbitrarily. We will assume that it is directed along Z axis
\cite{L1, L2, SU2}. The tensor $B_{\mu \nu}$ will be the following:
\[B_{\mu \nu}=\left( \begin{array}{cccc} 0 & -B & 0 & 0 \\ B & 0 & 0 & 0 \\ 0 & 0 & 0 & \pm B \\ 0 & 0 & \mp B & 0
\end{array}\right) \]
It can be also rewritten in the following form
%\begin{equation}
$$
\vec{B}=(B_{1},B_{2},B_{3})=(0,0,B).
$$
%\end{equation}

In this work we consider small perturbative field $A^{a}_{\mu}$ on
the background of the nonperturbative one. We have retained terms up
to the second order in $A^{a}_{\mu}$ \cite{L2}. In this
approximation the potential of the perturbative field

$$
V_{A}=\frac14 g^{2}\left\{(\vec A^{a})^{2}(\vec A^{b})^{2}-(\vec
A^{a} \vec A^{b})^{2}\right\}.
$$
is neglected.

If $A_{1}^{1}=q_{1}$, $A_{2}^{2}=q_{2}$ and the other components of
the perturbative Yang-Mills fields are equal to zero, the potential
of the model is:
%\begin{eqnarray}
%V=B^{2}+g^{2}\upsilon^{2} (q_{1}^{2}+q_{2}^{2})+g B q_{1} q_{2}+
%\\ + \frac18 g^{2} B^{2} \left\{q_{1}^{2}(x^{2}+z^{2}-t^{2})
%+q_{2}^{2}(y^{2}+z^{2}-t^{2})\right\};
%\end{eqnarray}

$$
V=B^{2}+\frac12 g^{2}\upsilon^{2} (q_{1}^{2}+q_{2}^{2})+g B q_{1}
q_{2}+ \eqno{(1)}
$$
$$ + \frac18 g^{2} B^{2} \left\{q_{1}^{2}(x^{2}+z^{2}-t^{2})
+q_{2}^{2}(y^{2}+z^{2}-t^{2})\right\},
$$

%V=B^{2} + \frac12 g^{2} \left\{\frac14
%B^{2}(x^{2}+y^{2}+z^{2}-t^{2})+ \upsilon^{2}\right\}
%(q_{1}^{2}+q_{2}^{2}) - \notag \\- \frac18 g^{2} B^{2} (y^{2}
%q_{1}^{2} + x^{2} q_{2}^{2})
% + g B q_{1} q_{2}.

where $x$,$y$,$z$ are the spatial coordinates and t - time.

\section{Stability of the model}

Stability of the model is investigated using well known technique
based on the Toda criterion of local instability \cite{Salasnich}.
The sign of the Gaussian curvature determines whether the system is
stable. The positive curvature indicates stable dynamics, negative -
chaotic.

The Gaussian curvature for potential (1) has the following form

$$
K_{G}= C_{1} C_{2} - g^{2} B^{2},\eqno{(2)}
$$
where
$$
C_{1}=g^{2}\left[\upsilon^{2}+\frac14
B^{2}(x^{2}+z^{2}-t^{2})\right],
$$
$$
C_{2}=g^{2}\left[\upsilon^{2}+\frac14
B^{2}(y^{2}+z^{2}-t^{2})\right].
$$
It can be seen from the last expressions that the stability of the
model depends only on the set of parameters and coordinates. The
model has three parameters and four coordinates: the self-coupling
constant g, the value of the nonperturbative field B, the value of
the Higgs field $\upsilon$, three spatial coordinates $x$,$y$,$z$
and the time $t$. Therefore one can regard the seven-dimensional
space of parameters and coordinates values. Every point in this
space corresponds to the particular set of the parameters and
coordinates. This set determines the stability of the model. Thus
one can associate the point in the regarded space with the stability
of the model. Therefore the stability and instability regions are
presented in this space. The expressions for bound surfaces in the
seven-dimensional space between the stability and instability
regions demonstrate the spacing of these regions. These surfaces can
be obtained as the manifolds where the Gaussian curvature (2) is
equal to zero.

Different values of spatial coordinates $x$,$y$ and $z$ correspond
to different types of the potential. Our model has a distinguished
point ($x=y=z=0$). If $x=y=0$ the potential changes along Z axis.
This type of potential will be investigated in subsection \ref{axe}.
In subsection \ref{point} we consider the potential in an arbitrary
point (1).

\subsection{Stability of the model along Z axis}\label{axe}

The potential on Z axis is obtained from (1) by setting $x$ and $y$
to zero:
$$
x=0, \quad y=0.
$$
Thus (1) transforms to the next form
$$
V=B^{2}+\frac12 g^{2}\left\{\upsilon^{2}+\frac14 B^{2} (z^{2}-t^{2})
\right\}(q_{1}^{2}+q_{2}^{2})+g B q_{1} q_{2}\eqno{(3)}
$$

In this case we have five parameters.The Gaussian curvature (2) is
$$
K_{G}= g^{4}\left[\upsilon^{2}+\frac14 B^{2}(z^{2}-t^{2})\right]^{2}
- g^{2} B^{2}.
$$
The roots of this expression correspond to the bound surfaces
between stability and instability regions and have the following
form:
$$
(t^{2})_{\pm}=4 \frac{\upsilon^{2}}{B^{2}}+z^{2}\pm 4 \frac{1}{g
B}.\eqno{(4)}
$$
These expressions describe two four-dimensional surfaces in
five-dimensional space of parameters and coordinates values.
Analysis of the expression for the Gaussian curvature showed that
the instability region is situated between these bounds.

The expression
$$
\Delta(t^{2})=8 \frac{1}{g B}\eqno{(5)}
$$
describes the width of the instability region.

\begin{figure}[h!]
     \leavevmode
\centering
\includegraphics[width=7cm]{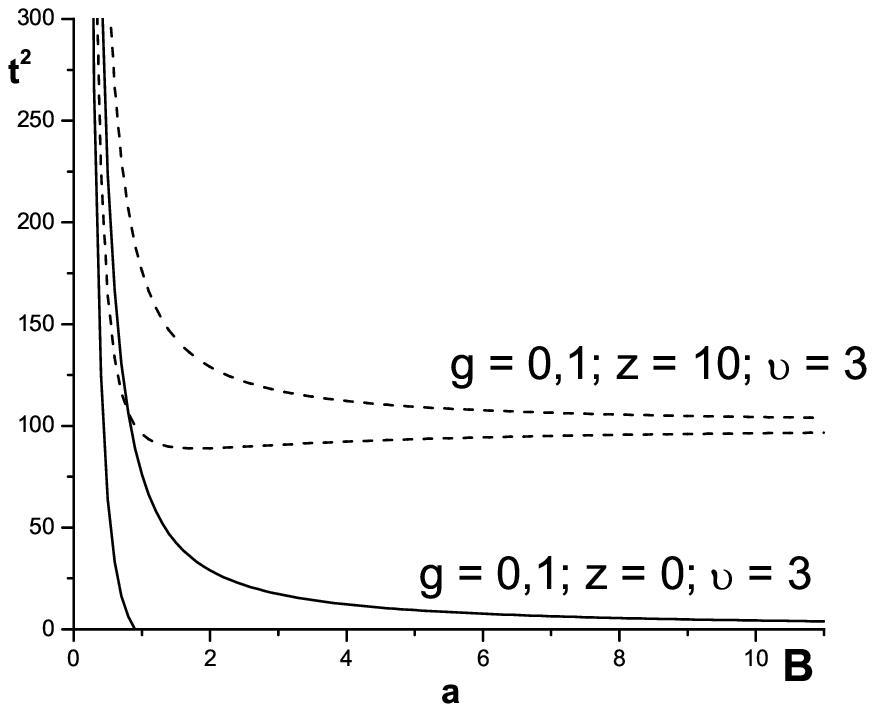}
\includegraphics[width=7cm]{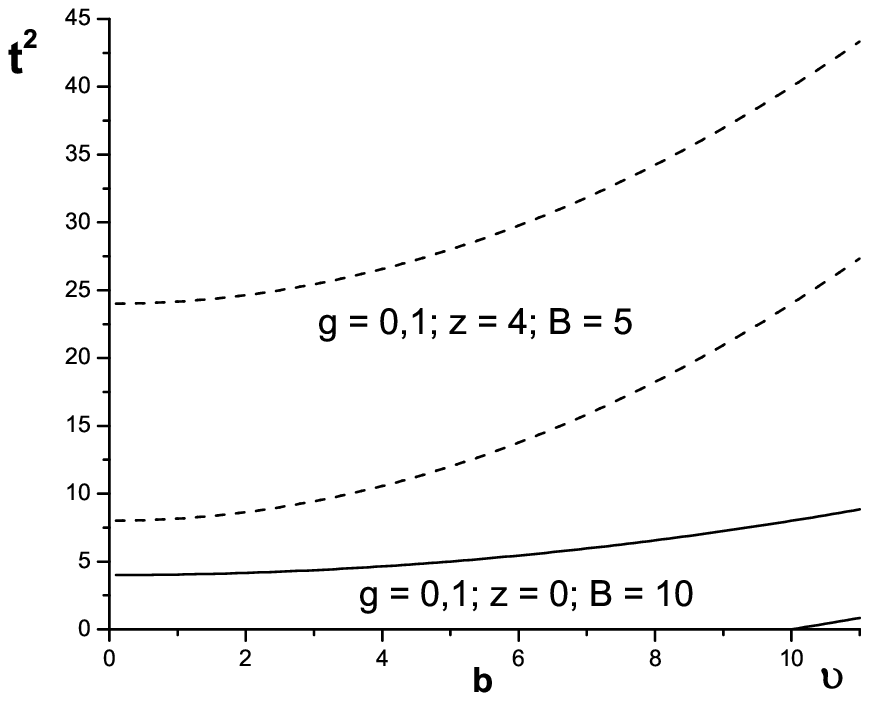}
\caption{The instability region for points on the axis Z (x = y = 0)
in plane \textbf{a)} ($t^{2}$ - B) and \textbf{b)} ($t^{2}$ -
$\upsilon$)} \label{GrBVg}
\end{figure}

It can be seen that in the case of the distinguished point (z=0) the
location of instability region depends on two parameters - the
values of the Higgs and nonperturbative fields. In the case of an
arbitrary point on Z axis the location of the  instability region is
shifted along time axis in comparison with the distinguished point
according to the expression (4). The width (5) of this region
remains unchanged. This case is illustrated on the first figure
(Fig.\ref{GrBVg}).

Two bound lines (4) are plotted on the figures. The instability
region is situated between them.

\begin{figure}[h!]
     \leavevmode
\centering
\includegraphics[width=7cm]{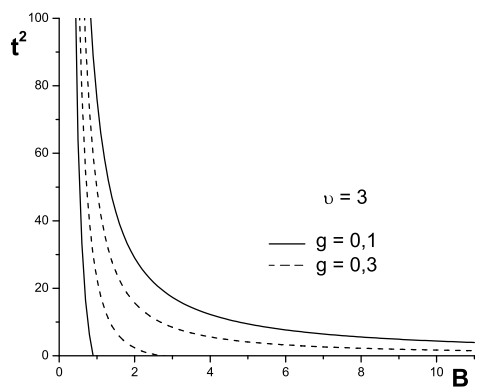}
\caption{The instability region in plane ($t^{2}$ - B) for
distinguished point (x = y = z = 0) } \label{GrBgg}
\end{figure}

The width of the instability region (5) depends on the value of the
nonperturbative field and the self-coupling constant. The first is
demonstrated on the first figure (Fig.\ref{GrBVg}). One can notice
that the self-coupling constant affects the width of the instability
region only (Fig.\ref{GrBgg}).

The potential energy (3) decreases with time. The density of energy
which corresponds to the chaos to order transition is:
$$
V_{+}=B^{2}-\frac12 g B (q_{1} - q_{2})^{2}.
$$
The sign '+' means that this transition occurs at time $(t)_{+}$.

\subsection{Stability of the model in the arbitrary
point} \label{point}

Now let us investigate the whole form of the potential (1). In this
case the potential depends on the point in the seven-dimensional
space of parameters and coordinates values. The Gaussian curvature
takes the general form (2). The expression for bound surfaces reads:
$$
(t^{2})_{\pm}=4 \frac{\upsilon^{2}}{B^{2}}+z^{2}+\frac12
x^{2}+\frac12 y^{2}\pm\eqno{(6)}
$$
$$
\pm4 \sqrt{\frac{1}{g^{2} B^{2}}+\frac{1}{2^{6}}(x^{2}-y^{2})^{2}}.
$$

The region under investigation is shifted along the axis T in
comparison with the case of the third axis. This shift is
proportional to the square of the coordinates $x$ and $y$.

\begin{figure}[h!]
     \leavevmode
\centering
\includegraphics[width=7cm]{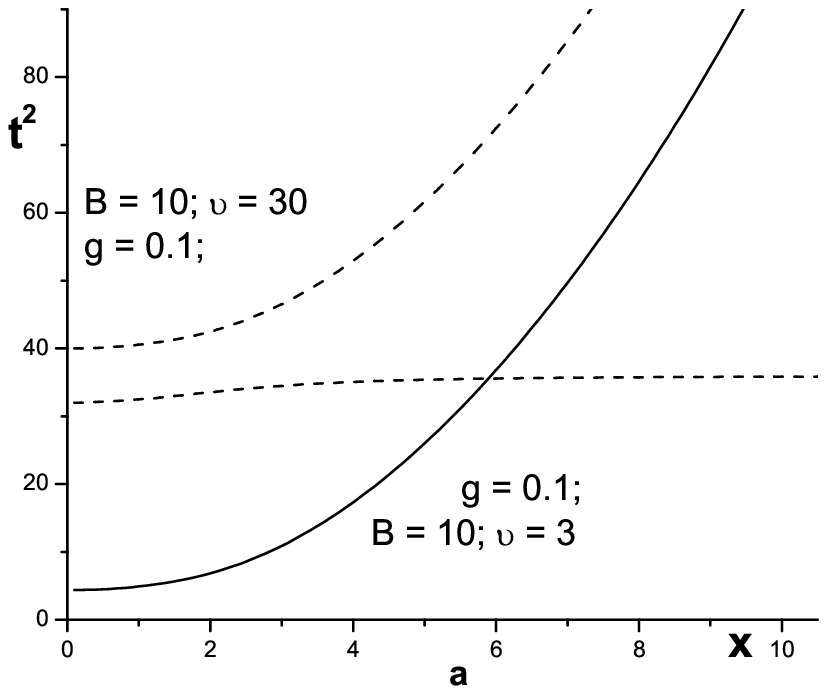}
\includegraphics[width=7cm]{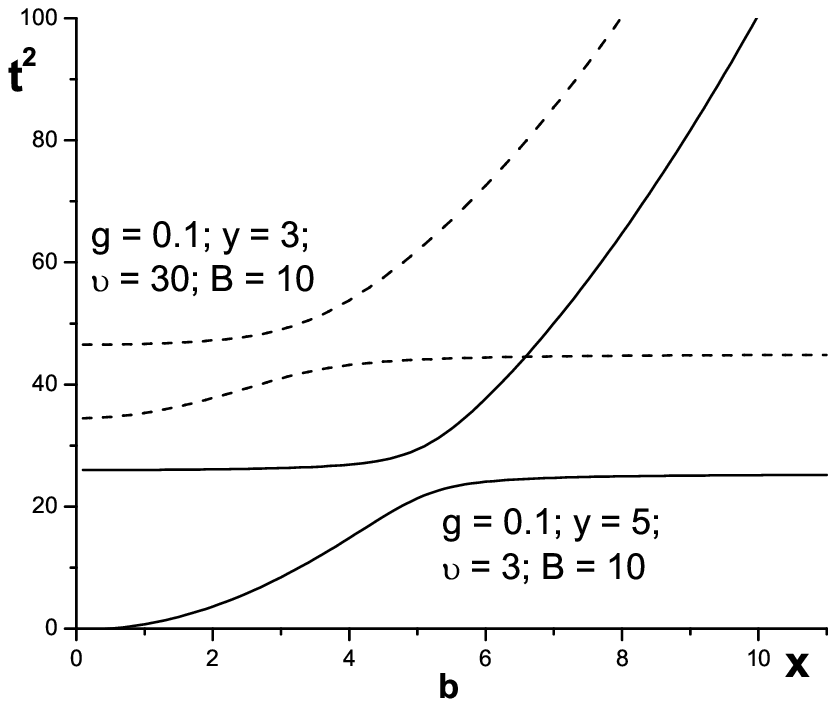}
\caption{The instability region in plane ($t^{2}$ - x) for points
\textbf{a)} on the axis X (y = z = 0) and \textbf{b)} on the line
which is parallel to axis X (z = 0) } \label{Gr05xg}
\end{figure}

Similar to the case of the third axis we obtain the expression for
the width of the instability region:
$$
\Delta(t^{2})=8 \sqrt{\frac{1}{g^{2}
B^{2}}+\frac{1}{2^{6}}(x^{2}-y^{2})^{2}}. \eqno{(7)}
$$
This expression is more complicated than the previous one. It
coincides with (5) when $x = y$. In other points it increases
strongly with the difference between $x$ and $y$ (Fig.\ref{Gr05xg}).

 Two bound lines (6) are plotted  on the figures. The
instability region is situated between them.

\begin{figure}[h!]
     \leavevmode
\centering
\includegraphics[width=7cm]{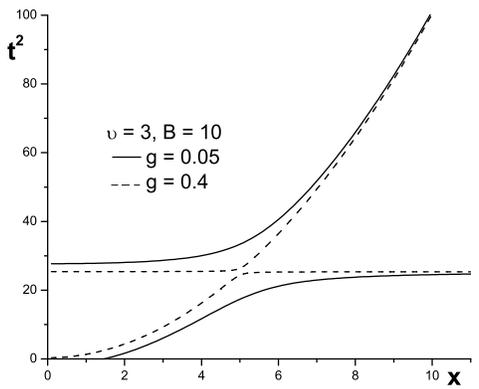}
\caption{The instability region in plane ($t^{2}$ - x) for points on
the line which is parallel to axis X (y = 5, z = 0)} \label{Gr5xgg}
\end{figure}

The coupling affects the width of the instability region when $x =
y$. Otherwise the width of the instability region mainly depends on
the difference between $x$  and $y$ (Fig.\ref{Gr5xgg}).

It can be seen from this figures that the type of stability is
instantaneously changed on the lines which are parallel to the axes
X and Y. It happens in the critical time which reads
$$
(t^{2})_{d}=4 \frac{\upsilon^{2}}{B^{2}} + z^{2} + u^{2},
$$
where $u=x$ or $u=y$ if the line is parallel to the axis X or Y
respectively. The direction of order-to-chaos transition depends on
the correlation between the coordinates $x$ and $y$. The example is
demonstrated on the Fig.\ref{Gr05xg}, where in the points with $x <
y$ chaos-order transition occurs, while the points with $y < x$
correspond to order-chaos transition. Discontinuously stability
change is'nt occurred on the lines which are non-parallel to the
axis X or Y.

\section*{Conclusions}

Classical dynamics of SU(2) model gauge field system with Higgs
field was considered in the homogeneous self-dual vacuum field. The
stability of the model was investigated based on the Toda criterion
of local instability. Analysis showed that the stability of the
model depends on the set of the parameters and coordinates. The
value of perturbative field does not influence on the model
stability in our approximation. Model has three parameters and four
coordinates: the self-coupling constant $g$, the value of the
nonperturbative field $B$, the value of the Higgs field $\upsilon$,
three spatial coordinates $x$, $y$, $z$ and time $t$. Therefore one
can associate the point in the seven dimensional space with the
stability of the model. The stability and instability regions were
detected. The analytical expressions for the bounds of the regions
were obtained for the points on Z axis $(x = y = 0)$ and for an
arbitrary point.

The moments of time of appearance and disappearance of the
instability region present in model with arbitrary set of parameters
and coordinates. The time of existence of the instability region
tends to infinity at small nonzero values of nonperturbative field
(Fig.\ref{GrBVg}, Fig.\ref{GrBgg}). This behavior could be connected
with confinement \cite{PLA, SU2, Simonov}.

\newpage
\section*{Figures}

\end{document}